\documentclass[10pt,twocolumn]{article}
\usepackage[a4paper,margin=0.8in]{geometry}
\usepackage{graphicx}\usepackage{amsmath,amssymb}\usepackage[T1]{fontenc}\usepackage{booktabs}
\usepackage{hyperref}\hypersetup{colorlinks=true,linkcolor=blue,citecolor=blue,urlcolor=blue}
\usepackage{caption}

\title{\bfseries\large Discovery of a gravitational arc candidate at $z_\mathrm{phot}\approx1.4$ in MACS\,J0308.9+2645 from a catalogue-based search of JWST imaging}
\author{Homer D\'avila Guti\'errez\\[2pt]
\small SKYCR.ORG, San Jos\'e, Costa Rica\\[1pt]
\small \href{https://orcid.org/0009-0002-5968-3646}{ORCID:\,0009-0002-5968-3646} \quad \texttt{cosmos@skycr.org}}
\date{Submitted to PASJ}
\begin{document}
\twocolumn[
\begin{@twocolumnfalse}
\maketitle
\begin{abstract}\noindent We report a previously uncatalogued gravitational arc candidate (A1) in the massive Reionization Lensing Cluster Survey (RELICS) cluster MACS J0308.9+2645 ($z_l$ = 0.356), identified in public JWST/NIRCam imaging from programme GO-5293 through a transparent catalogue-based selection of extended, elongated, tangentially aligned sources. A1 has axis ratio a/b $\approx$ 6.5, isophotal extent $\approx$ 5.1 arcsec, and lies 50.9 arcsec from the cluster X-ray centre, tangentially aligned to within 1.4$^\circ\pm$ 3.0$^\circ$ (the uncertainty dominated by the X-ray centroid). We show that the catalogue \texttt{aper\_total\_abmag} photometry, designed for unresolved sources, is invalid for this extended arc: the aperture captures only a few per cent of the isophotal flux and, because the aperture radius grows with wavelength, imposes $\approx$ 1.4 mag of spurious red colour that would drive a spurious high-redshift solution. Extended-source photometry --- isophotal magnitudes, fixed matched apertures, and curve-of-growth measurements, combined with archival HST/RELICS forced photometry in seven bands --- yields z $\approx$ 1.4 (plausible range 1.2--1.7). A significant detection in HST/ACS F435W independently excludes the spurious high-redshift solution, which would require a complete Lyman-limit dropout in that band. At z $\approx$ 1.4 the public Zitrin-LTM-Gauss lens model places A1 near the tangential critical curve in the single-image regime: full-map ray-tracing predicts no counter-images for z $\lesssim$ 2, consistent with the absence of any counterpart, whereas a z = 4.4 source at this position would require a counter-image of m $\approx$ 22.7 at an accessible location where none is observed. The most probable interpretation, is a singly lensed galaxy at z $\approx$ 1.4 behind the cluster. The arc's curvature is concave toward the cluster centre with radius of curvature $\approx$ 19 arcsec, but its statistical significance is marginal and method-dependent once correlated noise is properly accounted for; we document this explicitly. A fainter arclet (A2) is retained as a secondary candidate pending equivalent photometric reanalysis. Spectroscopic redshift determination and a dedicated lens model are required for confirmation.\end{abstract}
\vspace{0.3em}\noindent\textbf{Keywords:} gravitational lensing: strong --- galaxies: clusters: individual (MACS\,J0308.9+2645) --- galaxies: high-redshift
\vspace{1.2em}
\end{@twocolumnfalse}
]

\section{Introduction}
Strong gravitational lensing by galaxy clusters magnifies background galaxies and constrains the cluster mass distribution (Kneib \& Natarajan 2011). Identifying the arcs and multiple images that anchor lens models is therefore a central task, now transformed by wide-field, high-resolution imaging. A substantial methodological literature addresses lens finding, from convolutional neural networks trained on simulated and real data (Metcalf et al. 2019; Angora et al. 2023) to large-scale hybrid pipelines combining machine learning, citizen science, and expert vetting, as in the Euclid Strong Lens Discovery Engine (Euclid Collaboration: Lines et al. 2026; Euclid Collaboration: Bergamini et al. 2026). A complementary and deliberately simple approach remains useful for the specific morphology of cluster-scale tangential arcs: selecting, directly from a source catalogue, the extended and highly elongated objects oriented tangentially to the cluster centre, for subsequent vetting and modelling.

MACS J0308.9+2645 (hereafter MACS0308) is among the most massive Planck-selected clusters. Its published strong-lensing analysis, based on HST imaging within RELICS, identified multiple-image systems, measured a large effective Einstein radius ($\theta_E\gtrsim$ 30 arcsec at $z_s$ = 2), a projected mass of 2.5 $\pm$ 0.4 $\times10^{14}$ M$_\odot$ within the critical curves, and a bright multiply imaged z $\approx$ 6.2 source (Acebron et al. 2018). Recent JWST observations of this field have targeted that z $\approx$ 6.2 system, MACS0308zD1, the "Cosmic Spear" (Abdurro'uf et al. 2025). The cluster was observed with JWST/NIRCam on 2025 February 4 under programme GO-5293 (PI: X. Xu), now public at MAST. We adopt a Planck 2018 cosmology (Planck Collaboration et al. 2020); at $z_l$ = 0.356, 1 arcsec corresponds to 5.14 kpc. We document the \texttt{aper\_total} systematic explicitly (Sect.~3.2) because it constitutes a reproducible failure mode of general relevance to catalogue-based work on extended sources.

\section{JWST data and detection method}

\subsection{Data and source catalogue}
We use the public JWST/NIRCam Level-3 mosaics and the associated official source catalogues for GO-5293 in six bands (F115W, F150W, F200W, F250M, F300M, F410M). Candidate selection (Sect. 2.2) used the catalogue \texttt{aper\_total\_abmag} values and the astrometric and morphological parameters (ellipticity, semi-major axis, orientation) of the official F200W catalogue. As shown in Sect. 3.2, the \texttt{aper\_total} photometry is not valid for extended sources such as A1; all physical conclusions in this paper therefore rest on extended-source photometry: catalogue isophotal magnitudes, fixed matched-aperture measurements on the mosaics, curve-of-growth photometry, and forced photometry on the archival HST/RELICS 60 mas imaging (seven bands, ACS F435W/F606W/F814W and WFC3-IR F105W/F125W/F140W/F160W), after measuring and correcting a 0.37 arcsec astrometric offset between the pre-Gaia RELICS frame and the JWST frame. The cluster X-ray centre is from Cavagnolo et al. (2008).

\subsection{Candidate selection and validation}
From the F200W catalogue we select tangential-arc candidates with transparent criteria. For each extended source (catalogue flag \texttt{is\_extended} ) brighter than AB = 28 we require: ellipticity $\epsilon\geq$ 0.55; semi-major axis $\geq$ 6 pixels ($\approx$ 0.19 arcsec); and projected separation from the cluster centre between 3 and 70 arcsec. For each surviving source we compute the position angle of the centre-to-source vector and the local tangential direction, and retain sources whose catalogue orientation lies within 35$^\circ$ of tangential. Candidates are ranked by a score S = $\epsilon(a/10\,\mathrm{px})^{1/2}$ (1 - $\Delta\theta$/120$^\circ$), where a is the semi-major axis and $\Delta\theta$ the misalignment from tangential; the full definition and the injection tests are provided in the public repository (Sect. Data availability).

The method's behaviour was characterised in two complementary ways. First, to test morphological false-positive control, we applied the selection across 54 public JWST/NIRCam fields --- open quasar fields from the ASPIRE and EIGER programmes, galaxy clusters, and calibration fields --- evaluating 1591 extended sources (Table~\ref{tab:campaign}). High-significance (> 8$\sigma$) isophote-residual sources are dominated by astrophysical false positives (clumpy edge-on discs, barred and two-armed spirals, radial features), removed by morphological vetting and visual inspection. Second, completeness was estimated by injecting synthetic arcs of known brightness into real cluster cutouts and recovering them with the same pipeline. Across the full campaign, only two sources, both in MACS0308, survived as strong tangential-arc candidates: A1 and A2.

\begin{table*}[tbp]\centering\begin{tabular}{lrrrr}\toprule Field type & $N$ fields & $N$ galaxies & High-sig ($>8\sigma$) & Strong arc cand. \\ \midrule ASPIRE quasar fields & 25 & 760 & 67 & 0 \\ EIGER quasar fields & 9 & 285 & 36 & 0 \\ Galaxy clusters & 10 & 491 & 37 & 2 (A1, A2) \\ Calibration/other & 10 & 55 & 3 & 0 \\ \midrule \textbf{Total} & \textbf{54} & \textbf{1591} & \textbf{143} & \textbf{2} \\ \bottomrule \end{tabular}\caption{Summary of the 54-field detection campaign used to characterise morphological false-positive control. For each field type we list the number of fields, the number of extended sources evaluated, those with $>8\sigma$ isophote residuals, and the surviving strong tangential-arc candidates. Only MACS0308 yields candidates (A1, A2).}\label{tab:campaign}\end{table*}

\section{Results}

\subsection{Astrometry, morphology, and curvature}\label{sec:astro}
\begin{figure*}[tbp]\centering\includegraphics[width=0.92\textwidth]{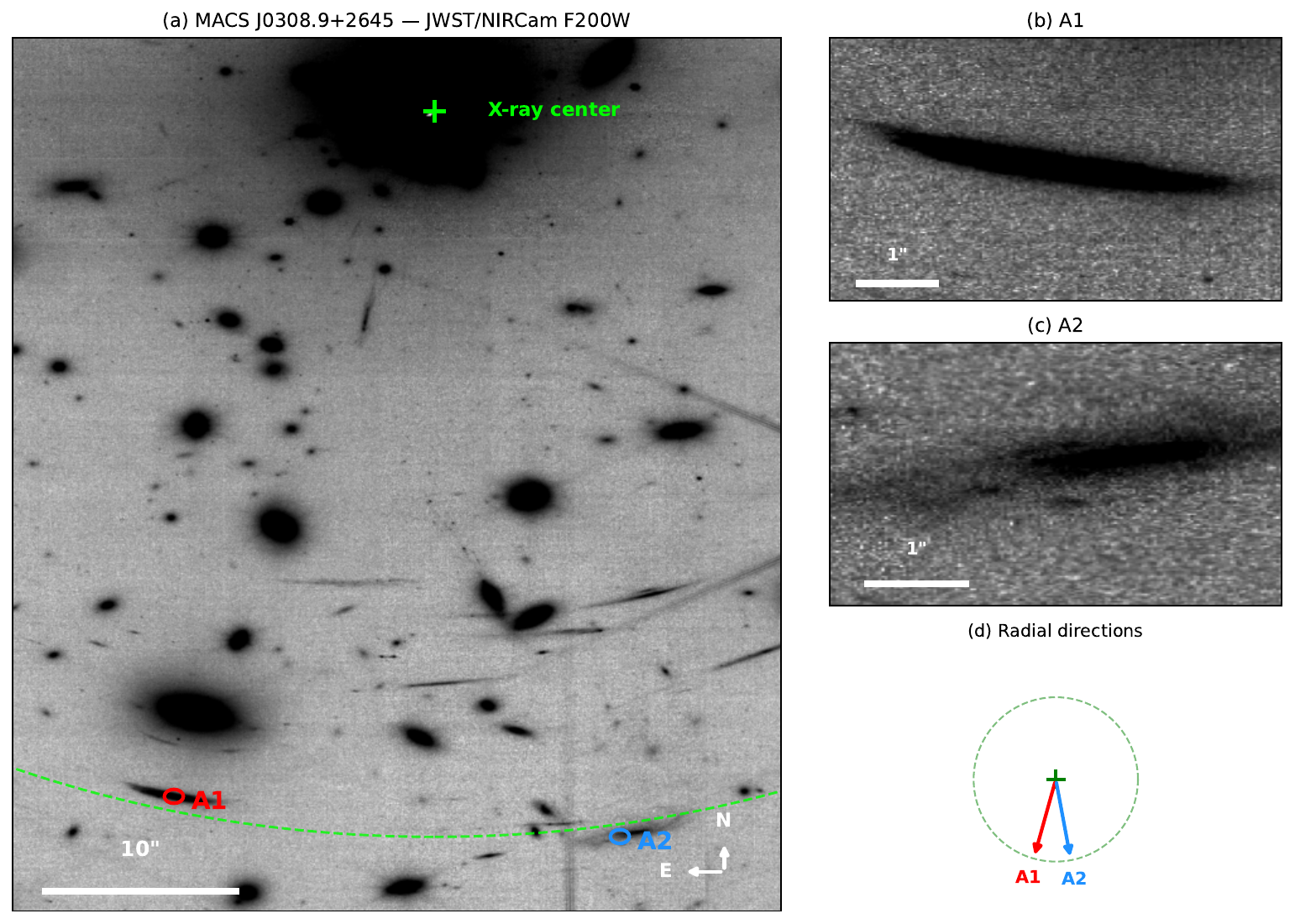}\caption{(a) JWST/NIRCam F200W image of MACS0308 with A1 (red), A2 (blue), and the X-ray centre (green cross); dashed circle marks the common radius ($\approx51.9^{\prime\prime}$). (b) A1. (c) A2. (d) Radial directions from the X-ray centre to A1 and A2.}\label{fig:field}\end{figure*}

The highest-ranked previously unreported candidate, A1 (catalogue label 244), lies at R.A. = 03:08:52.64, Dec. = +26:45:11.8 (J2000); the field configuration and close-ups are shown in Fig.~\ref{fig:field}, and the astrometric and morphological parameters of both candidates are collected in Table~\ref{tab:astro}. Its official catalogue length is 3.9 arcsec, while the 3$\sigma$ isophotal mask traces a lower-surface-brightness chord of 5.1 arcsec. It has catalogue axis ratio a/b $\approx$ 6.5 ($\epsilon$ = 0.85) and lies at a projected radius of 50.9 arcsec from the X-ray centre. Blind, independent source extraction on the JWST F200W and HST/ACS F435W mosaics --- two telescopes, two epochs, two pipelines --- recovers the same object with centroids agreeing to 0.21 arcsec, sky position angles to 3.4$^\circ$, and semi-major axes to 3 per cent, excluding an instrumental origin.

A1 is tangentially aligned to within $\Delta\theta_\mathrm{tan}$ = 1.4$^\circ\pm$ 3.0$^\circ$. The uncertainty budget comprises the dispersion among three independent position-angle measurements (1.9$^\circ$) and, dominantly, the unpublished uncertainty of the X-ray centroid: at r = 50.9 arcsec, a centroid error of e arcsec rotates the tangential direction by arctan(e/r), i.e. 2.3$^\circ$ for e = 2 arcsec. The alignment is therefore consistent with exactly tangential but does not by itself exclude misalignments of a few degrees.

\begin{figure*}[tbp]\centering\includegraphics[width=0.82\textwidth]{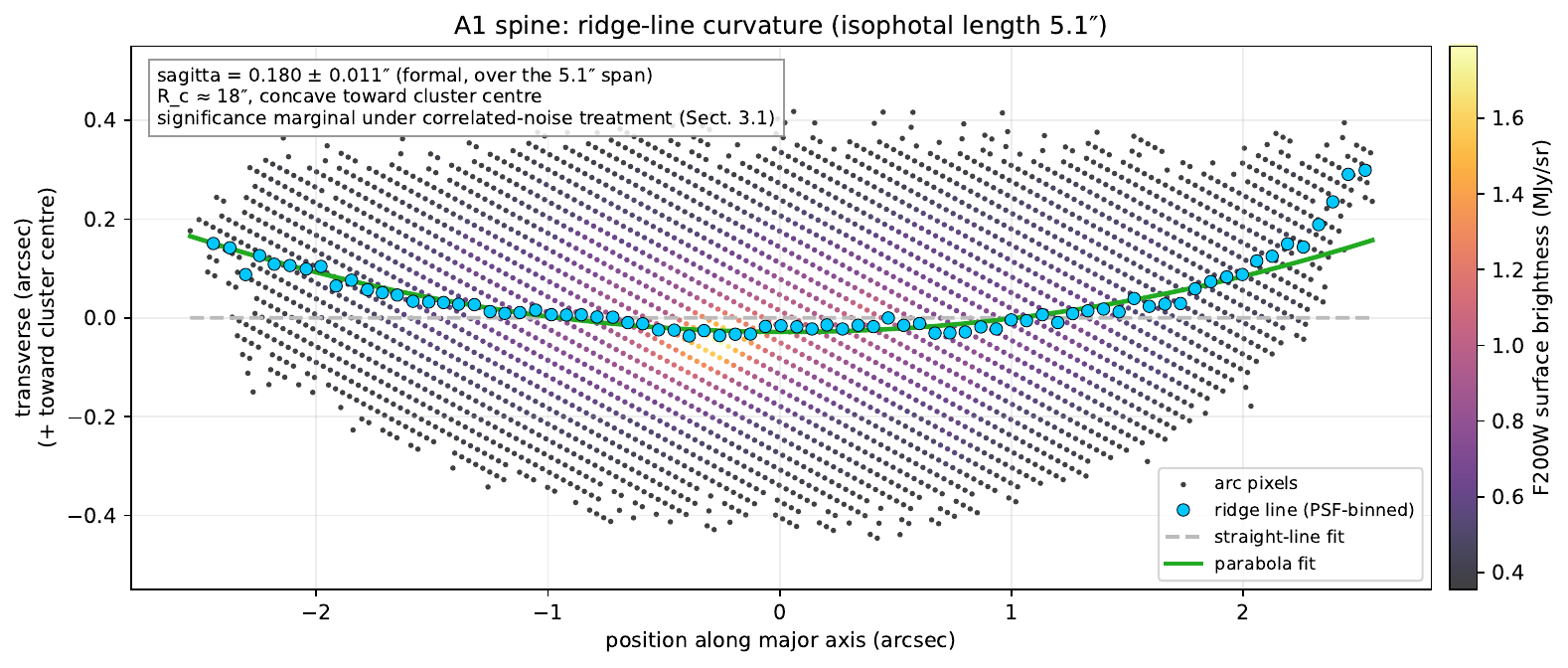}\caption{Curvature of A1. The surface-brightness-weighted ridge line departs from a straight-line fit and follows a parabola with formal sagitta $0.180\pm0.011^{\prime\prime}$ over the declared $5.1^{\prime\prime}$ isophotal span, concave toward the cluster centre, $R_c\approx18^{\prime\prime}$. The formal error underestimates the true uncertainty; significance is marginal once correlated noise is accounted for (Sect.~\ref{sec:astro}).}\label{fig:curv}\end{figure*}

Fitting the surface-brightness-weighted ridge line (Fig.~\ref{fig:curv}) over the full 5.12 arcsec isophotal span with a straight line and a parabola gives a formal sagitta of 0.180 $\pm$ 0.011 arcsec, concave toward the cluster centre, with a circular radius of curvature $R_c\approx$ 18 arcsec; $R_c$ is stable to within a factor of 1.5 against the choice of span, whereas the sagitta scales as the span squared and is therefore only meaningful with the span declared. We caution that a naive bootstrap of the binned ridge points would formally reject zero curvature at high significance, but such a bootstrap treats points separated by less than the PSF as independent and is not valid: injection of intrinsically straight synthetic arcs of A1's brightness and length into empty regions of the mosaic shows that the empirical dispersion of the recovered curvature coefficient exceeds the naive bootstrap error by more than an order of magnitude. A forward model comparison on the original pixels --- straight versus curved ridge with identical freedom in centroid, position angle, width, longitudinal profile, and local background, under noise whitened with the empirically measured local power spectrum and calibrated by parametric bootstrap --- prefers curvature toward the centre with best-fitting $R_c\approx$ 18 arcsec, but at marginal and mask-dependent significance (empirical p between $\approx$ 0.03 and $\approx$ 0.5 across reasonable analysis choices). We therefore report the curvature as suggestive but not established; the direction (concave toward the cluster) is the aspect most consistently recovered.

A fainter arclet, A2 (catalogue label 218 in F200W), lies at R.A. = 03:08:51.97, Dec. = +26:45:33.6, at 52.7 arcsec, also roughly tangentially aligned (position angle within 2.3 arcdeg of tangential); a fainter neighbouring catalogue entry lies within 0.9 arcsec. We report A2 as a secondary candidate only.

\begin{table*}[tbp]\centering\begin{tabular}{lllllllll}\toprule ID & R.A. (J2000) & Dec. (J2000) & $r_X$ & Length & $a/b$ & $\epsilon$ & $\Delta\theta_\mathrm{tan}$ & $z$ \\ \midrule A1 & 03:08:52.64 & $+$26:45:11.8 & $50.9^{\prime\prime}$ & $3.9^{\prime\prime}$ & 6.5 & 0.85 & $1.4^\circ\!\pm\!3.0^\circ$ & $\approx1.4$ \\ A2 & 03:08:51.97 & $+$26:45:33.6 & $52.7^{\prime\prime}$ & $\sim2.5^{\prime\prime}$ & 3.9 & 0.73 & $2.3^\circ$ & pending \\ \bottomrule \end{tabular}\caption{Astrometry, morphology, and adopted redshift of the two candidates. Columns: identifier, J2000 coordinates, projected radius from the X-ray centre, arc length, axis ratio, ellipticity, misalignment from the tangential direction, and adopted redshift. A1 $=$ catalogue label 244; A2 $=$ catalogue label 218 (F200W). The A2 redshift is pending reanalysis with extended-source photometry (Sect.~\ref{sec:phot}).}\label{tab:astro}\end{table*}

\subsection{Photometry and photometric redshifts}\label{sec:phot}
A naive EAZY fit to the six-band catalogue \texttt{aper\_total\_abmag} photometry of A1 yields an apparently well-constrained high-redshift solution (z $\approx$ 4.4--5, $P(z>3)\approx$ 1). That solution is an artefact of the photometric method, which we document because it is a reproducible failure mode for any catalogue-based analysis of extended sources. The JWST pipeline documentation specifies that \texttt{aper\_total\_abmag} is intended for unresolved sources; A1 is flagged \texttt{is\_extended} . For A1 in F200W, \texttt{aper\_total\_abmag} = 24.53 while the same catalogue's isophotal\_abmag = 20.87: the aperture magnitude misses a factor of $\approx$ 29 (3.7 mag) of the flux, capturing $\approx$ 3 per cent of the isophotal light. The six-band isophotal photometry is given in Table~\ref{tab:phot}, alongside the \texttt{aper\_total} values to document the artefact. Moreover, the effective aperture radius grows with wavelength (r70 from 0.103 arcsec at F115W to 0.202 arcsec at F410M), so the captured fraction of the 5 arcsec arc rises from $\approx$ 3 to $\approx$ 10 per cent across the six bands, imposing $\approx$ 1.4 mag of purely geometric red colour. Fitting that spurious red continuum is what produces the apparent high-redshift solution.

\begin{figure*}[tbp]\centering\includegraphics[width=0.92\textwidth]{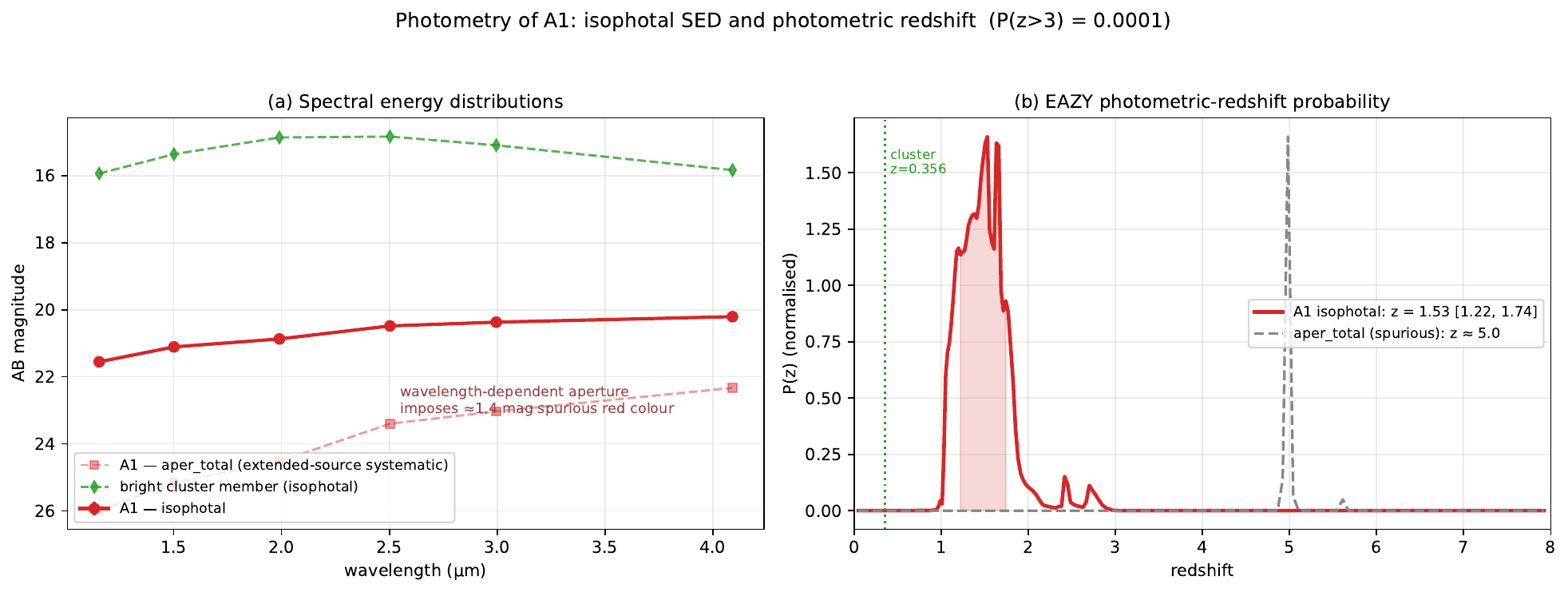}\caption{(a) Isophotal SED of A1 (red solid) vs.\ the invalid \texttt{aper\_total} SED (red dashed) and a bright cluster member (green). (b) EAZY $P(z)$: $z=1.53$ [1.22, 1.74], $P(z>3)=0.0001$; grey dashed shows the spurious \texttt{aper\_total} solution.}\label{fig:sed}\end{figure*}

Extended-source photometry was obtained by four mutually independent methods: per-band isophotal magnitudes within a fixed segmentation mask; fixed matched circular apertures; PSF-matched fixed-mask photometry; and curve-of-growth measurements. All four agree, and EAZY fits to the resulting six-band photometry yield z = 1.53 with a 16--84 per cent interval of [1.22, 1.74] and $P(z>3)$ = 0.0001 (Fig.~\ref{fig:sed}). A joint thirteen-band fit adding forced photometry on the archival HST/RELICS imaging (seven bands, astrometric offset of 0.37 arcsec measured from 153 cross-matched sources and corrected) gives a best solution at z $\approx$ 1.2--1.4; we adopt z $\approx$ 1.4 with a plausible range of 1.2--1.7. An independent consistency check requires no photometric calibration at all: A1 is clearly detected in HST/ACS F435W (integrated $\approx$ 12$\sigma$ after correlated-noise correction, with the arc's own elongated morphology). At z = 4.4 the Lyman limit (rest 912 \AA{}) falls at 4925 \AA{}, redward of the F435W pivot wavelength (4319 \AA{}), so the source would be a complete dropout in that band; the detection therefore excludes the high-redshift solution independently of aperture corrections, template sets, or fitting codes.

An \texttt{aper\_total} -based fit for A2 ($z_\mathrm{phot}\approx$ 3.0) carries the same systematic error, compounded by its faintness; we regard its redshift as undetermined pending equivalent extended-source photometry.

\begin{table}[tbp]\centering\begin{tabular}{lccc}\toprule Band & isophotal & \texttt{aper\_total} & $\Delta$ \\ & (AB) & (AB) & (mag) \\ \midrule F115W & $21.55$ & 26.00 & $+4.45$ \\ F150W & $21.11$ & 25.16 & $+4.05$ \\ F200W & $20.87$ & 24.53 & $+3.66$ \\ F250M & $20.48$ & 23.40 & $+2.92$ \\ F300M & $20.37$ & 23.03 & $+2.66$ \\ F410M & $20.21$ & 22.33 & $+2.12$ \\ \bottomrule \end{tabular}\caption{Six-band isophotal photometry of A1 from the official JWST L3 catalogue, with the \texttt{aper\_total} values listed to document the extended-source systematic. Statistical errors are $<0.01$ mag; a 0.05 mag systematic floor is adopted. The wavelength-dependent deficit ($+4.45$ to $+2.12$ mag) is the origin of the spurious red colour.}\label{tab:phot}\end{table}

\subsection{Position relative to the critical curve}
\begin{figure*}[tbp]\centering\includegraphics[width=0.62\textwidth]{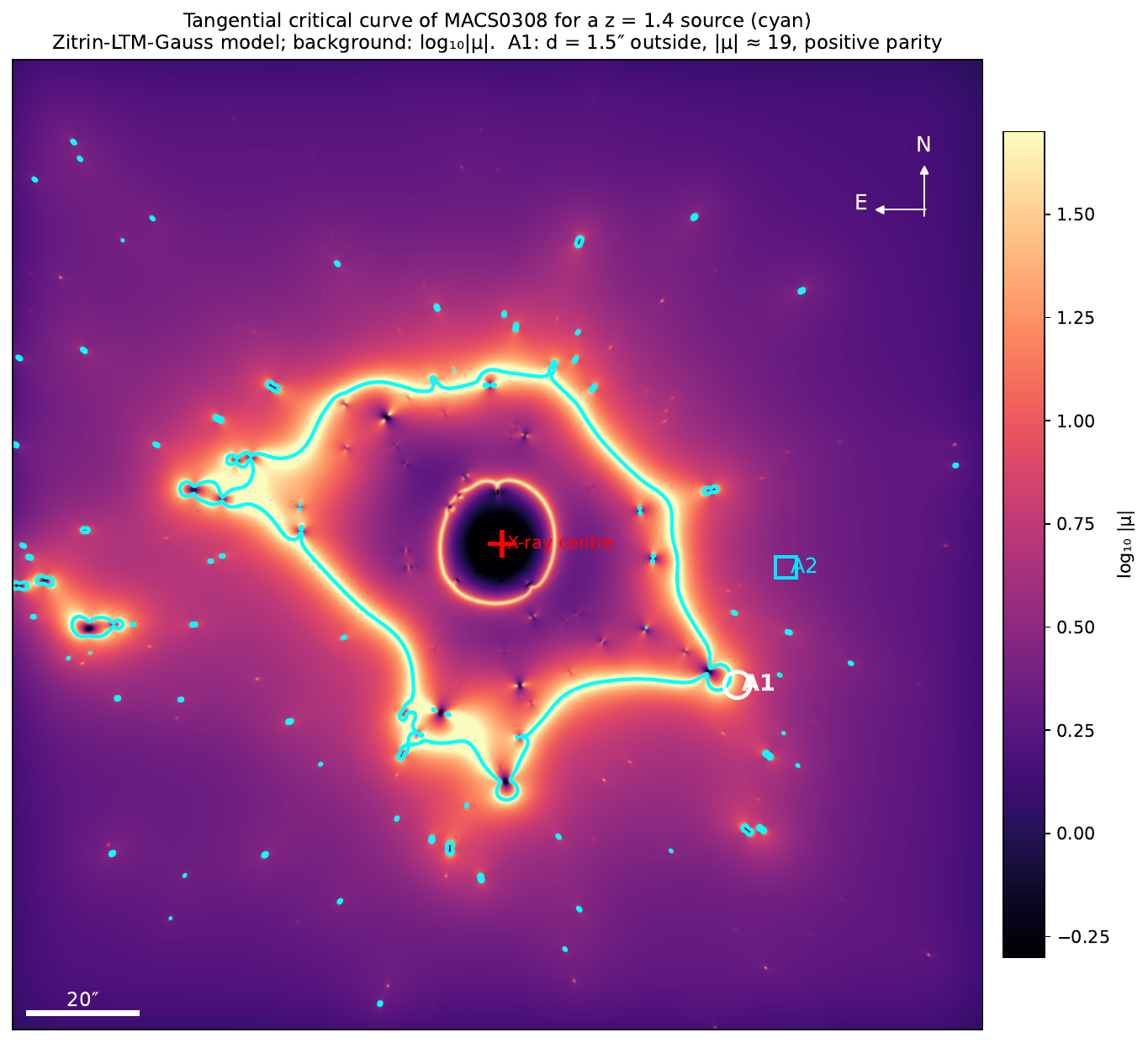}\caption{Tangential critical curve of MACS0308 for a $z=1.4$ source (cyan; Zitrin-LTM-Gauss) over $\log_{10}|\mu|$. A1 (white) lies $1.6^{\prime\prime}$ outside, formal $|\mu|\approx19$, positive parity; at $z_s=1.2$, $2.7^{\prime\prime}$, $|\mu|\approx10$. A2 (cyan square), X-ray centre (red cross).}\label{fig:crit}\end{figure*}

Using the public Zitrin-LTM-Gauss lens model of MACS0308 (Acebron et al. 2018; constructed from HST data independently of this work), we computed the convergence and shear at A1 from the public $\kappa$ and $\gamma$ maps, scaled from the source-at-infinity normalisation by the factor Dls/Ds appropriate to each source redshift, and additionally ray-traced the full public deflection maps of the best-fitting model. The lensing configuration at the adopted redshift is shown in Fig.~\ref{fig:crit}. For $z_s$ = 1.2--1.4, A1 lies $\approx$ 1.5--3 arcsec outside the tangential critical curve (2.7 arcsec at $z_s$ = 1.2, 1.6 arcsec at $z_s$ = 1.4), with formal point magnification |$\mu$| $\approx$ 10--20 and positive parity, in the single-image regime; given the proximity to the critical curve, the point magnification is sensitive to the exact source redshift and a robust value requires an extended-source treatment. Back-projecting the observed ridge line of A1 to the source plane and mapping the entire field returns no additional image solutions for $z_s\lesssim$ 2. The absence of a counter-image is therefore the model's prediction at the adopted redshift, not a deficiency of the candidate. Conversely, for a hypothetical $z_s$ = 4.4 source at the position of A1 the same model predicts a counter-image of m $\approx$ 22.7 at an accessible position $\approx$ 22 arcsec from the X-ray centre, where the deepest catalogue entry within 1 arcsec is 3.6 mag fainter and morphologically inconsistent; this constitutes an independent, photometry-free argument against the high-redshift solution of Sect. 3.2. We note that at $z_s\approx$ 1.4 the source-plane reconstruction of A1 is also most compact ($\approx$ 0.1--0.3 arcsec), as expected if the elongation is dominated by lensing shear at approximately this redshift.

\subsection{Novelty}
Neither A1 nor A2 appears in the multiple-image inventory of Acebron et al. (2018). Neither source is present within 5 arcsec in SIMBAD, NED, or VizieR. Both are spatially distinct from the known z $\approx$ 6.2 lensed source in this field (MACS0308zD1, the "Cosmic Spear"), whose brightest image lies 16 arcsec away, and from its counter-image reported by Abdurro'uf et al. (2025), which lies 71 arcsec away.

\section{Lensing interpretation and limitations}
The evidence that A1 is a lensed image of a background galaxy comprises: a robust redshift placing it well behind the lens (z $\approx$ 1.4 versus $z_l$ = 0.356); extreme elongation (a/b $\approx$ 6.5) tangentially aligned with the cluster within the measurement uncertainty; position near the tangential critical curve for its redshift, in a |$\mu$| $\approx$ 10 region of one of the most massive known clusters; a source-plane reconstruction that is most compact near the adopted redshift; suggestive (though not statistically established) curvature concave toward the cluster centre with $R_c\approx$ 19 arcsec; and consistency between the single-image prediction of the public lens model at z $\approx$ 1.4 and the observed absence of counterparts.

If A1 traces the cluster potential at its position, the implied projected mass within its radius is M(< 50.9 arcsec) $\approx$ 5 $\times10^{14}$ M$_\odot$ for $z_s\approx$ 1.4 (Planck 2018 cosmology; the RELICS maps adopt H0 = 70, a 3 per cent effect on the mass). We report this as the mass implied by the arc position under circular symmetry, conditional on the adopted redshift, not as an independent measurement.

A1 remains a candidate. The principal non-lensing alternative is an intrinsically elongated (edge-on) disc galaxy at z $\approx$ 1.4 projected behind the cluster with fortuitous tangential orientation: the curvature significance is marginal (Sect. 3.1), so morphology alone cannot exclude it. Distinguishing the two requires a dedicated lens model of the JWST-era cluster (which would predict the shear at A1's exact position) and, definitively, spatially resolved spectroscopy: a lensed image's velocity field, delensed with the cluster model, must be consistent with a normal rotating galaxy of its mass, whereas an unlensed disc requires no such correction. Spectroscopy would simultaneously fix the source redshift, on which the magnification and mass scale.

\section{Conclusions}
We have reported A1, a previously uncatalogued gravitational arc candidate in MACS0308, identified in public JWST imaging with a transparent, reproducible catalogue-based selection, together with a secondary candidate A2. Multi-instrument extended-source photometry places A1 at z $\approx$ 1.4 (range 1.2--1.7), firmly behind the cluster; an HST/ACS F435W detection independently excludes the spurious high-redshift solution that point-source aperture photometry would suggest, a failure mode we document explicitly. At this redshift the public HST-era lens model predicts A1 to be a singly lensed image with |$\mu$| $\approx$ 10 and no counter-images, consistent with observation. The immediate next steps are spectroscopic confirmation and analysis with a dedicated JWST-era lens model of the cluster.

\section*{Acknowledgements}
Based on observations made with the NASA/ESA/CSA James Webb Space Telescope, obtained from MAST at STScI, associated with programme GO-5293, and on archival observations from the RELICS Hubble Treasury Program. This work uses the RELICS lens model of MACS0308 (Zitrin-LTM-Gauss). This research used the NASA/IPAC Extragalactic Database (NED) and the SIMBAD and VizieR databases (CDS, Strasbourg). Photometric redshifts were computed with EAZY. Generative AI tools were used to assist in the development of the detection pipeline and in the post-submission verification analyses; all numerical results and conclusions were independently verified by the author, who assumes full responsibility for this work.

\section*{Data availability}
The detection pipeline and derived measurements are available on Zenodo (\href{https://doi.org/10.5281/zenodo.21233185).}{doi:10.5281/zenodo.21233185).} The JWST data are public at MAST (\href{https://doi.org/10.17909/wj3v-0a65).}{doi:10.17909/wj3v-0a65).}

\end{document}